\documentclass[twocolumn,iop]{emulateapj}
\usepackage{graphicx}
\usepackage{natbib}
\slugcomment{Received 2015 February 4; accepted 2015 April 15; published 2015 June 5}
\begin{document}
\title{Numerically Predicted Indirect Signatures of Terrestrial Planet Formation}
\author{Zo\"e M. Leinhardt, Jack Dobinson, Philip J. Carter, Stefan Lines}
\affil{School of Physics, University of Bristol, HH Wills Physics Laboratory, Tyndall Avenue, Bristol BS8 1TL, UK}

\begin{abstract}
The intermediate phases of planet formation are not directly observable due to lack of emission from planetesimals. Planet formation is, however, a dynamically active process resulting in collisions between the evolving planetesimals and the production of dust. Thus, indirect observation of planet formation may indeed be possible in the near future. In this paper we present synthetic observations based on numerical N-body simulations of the intermediate phase of planet formation including a state-of-the-art collision model, EDACM, which allows multiple collision outcomes, such as, accretion, erosion, and bouncing events. We show that the formation of planetary embryos may be indirectly observable by a fully functioning ALMA telescope if the surface area involved in planetesimal evolution is sufficiently large and/or the amount of dust produced in the collisions is sufficiently high in mass.
\end{abstract}


\section{Introduction}

Planet formation produces a broad range of planetary outcomes ranging from hot Jupiters around main-sequence solar-type stars \citep{Mayor:1995,Marcy:1997} to lunar-mass planets around evolved degenerate neutron stars \citep{Wolszczan:1992}. Planets appear to be a common by-product of star formation with at least one in every four adult stars found to have at least one planet \citep{Petigura:2013}. Recent advances in observational techniques and hardware have led to a huge increase in the number and diversity of extrasolar planets \citep{Lissauer:2014} and unprecedented resolution and detail in young protoplanetary disks. However, these observations represent snapshots or still frames of the beginning and end of a dynamic evolving story of planet formation giving us just the briefest look behind the scenes. 

The intermediate stages of planet formation are difficult to observe directly for a number of reasons: first, planetesimals, the building blocks of planets, have a small surface area to mass ratio thus they are difficult to observe in reflected light from the star, second, planetesimals are relatively low-mass objects, thus, they do not produce significant thermal energy themselves nor do they produce any significant gravitational effect on the central star; third, young stars are notoriously noisy at a number of wavelengths making searches for faint objects near the central star more difficult. In the core accretion model of planet formation planetesimals evolve into planets and embryos via collisions with one another. All collisions whether accretion dominated or erosive result in the production of some dust and dust has a high surface area to mass ratio making it highly detectible at infra-red and sub-mm wavelengths. Thus, although direct observation of planetesimal evolution and planet formation may not be possible in the near future it may be more prudent to search for indirect signatures by looking for collisionally generated dust. 

In this work we present numerical simulations of terrestrial planet formation including a state-of-the-art planetesimal collision model, which can describe a range of collision outcomes from merging to catastrophic disruption. Using the collisions from the evolving planetesimal population we construct synthetic dust images and find that terrestrial planet formation may just be observable with a fully operational ALMA telescope.

\section{Numerical Method}

All simulations were carried out using the parallelised $N$-body gravity code PKDGRAV \citep{Stadel:2001,Richardson:2000,Leinhardt:2005}. PKDGRAV is a second order leap-frog integrator with a hierarchical tree to efficiently calculate gravitational interactions between large numbers of particles.

\subsection{Planetesimal Disk Model}

All simulations presented here are of modest resolution ($N = 10^4$) in order to compare directly with previous work \citep{Kokubo:2002,Leinhardt:2005}. The simulations begin with equal-sized particles arranged in an annulus from 0.5 to 1.5 AU around a single central stellar potential of one solar mass. The particles are distributed with a power-law surface density distribution $\Sigma = \Sigma_1(a/\mathrm{AU})^{-\alpha}$, where $\alpha = 3/2$. The total mass in solid material is $2.8 M_\oplus$ similar to the mass derived from the minimum mass solar nebula. 

Planetesimal particles were radially expanded by a factor $f$ in order to accelerate the planetesimal evolution. The inflation of the planetesimal particles reduces the evolution timescale by approximately $f^2$ \citep{Kokubo:2002}. The specific value of the expansion factor ($f = 6$) was chosen to be consistent with previous work. It has been shown analytically in \citet{Kokubo:1996} that $f \sim 6$ should not significantly impact the growth phase of planetesimals as long as gravitational scattering is relatively unimportant and the planetesimals are evolving in a simple system with a single central potential and no giant planet perturbers. For simplicity nebular gas is neglected in this work. Planetesimals will be negligibly effected by this simplification (they are large and would suffer little aerodynamic drag though collisionally generated debris would be more substantially affected).\footnote{ We are currently working on simulations with a lower expansion factor and including nebular gas that would allow us to complete more diverse simulations that evolved planetesimals to a later state (see Carter et al. in prep).}

\subsection{Planetesimal Collision Model}

In this paper we compare terrestrial planet formation simulations using three collision models: perfect merging, RUBBLE \citep{Leinhardt:2005}, and our new collision model EDACM \citep{Leinhardt:2012}. In both the RUBBLE and EDACM collision models two timesteps are used within the simulation to allow particle collisions to be accurately resolved but keep the simulation as efficient as possible. The major ``orbital" timestep is determined by the orbital dynamical time ($\sim 1$ yr at 1 AU), however, the ``collisional" timestep is determined by the planetesimal dynamical time ($\sim 1/\sqrt{G \rho}$) which is of order hours ($\rho = 0.00925\, \mathrm{g\,cm}^{-3}$ for $f=6$). Therefore, in all EDACM and RUBBLE simulations all planetesimals were initially placed on the major timestep (0.01 yr), when a collision was detected the impactors were demoted to the minor collisional timestep ($1.5\times10^{-4}$ yr) and the gravity is calculated about 64 times for the colliding particles and once for all other particles \citep[for details see][]{Leinhardt:2005,Bonsor:2015}. 

\subsubsection{Perfect Merging}

The simplest collision model used in these simulations is perfect merging. In our implementation of this model two colliding planetesimals merge on impact assuming a perfectly inelastic collision if the impact speed is less than or equal to the mutual particle escape speed regardless of impact angle or mass ratio. If the impact speed is greater than the particle escape speed the impactors bounce inelastically with a normal coefficient of restitution of 0.8. This collision model is the most common model used in previous $N$-body simulations \citep[for example,][]{Kokubo:1998,Kokubo:2000,Kokubo:2002}. 

\subsubsection{RUBBLE} \label{sec:rubble}

In addition to perfect merging we compare our new collision model to RUBBLE the collision model used in \citet{Leinhardt:2005}. RUBBLE is nominally more accurate than perfect merging and allows for a variety of collision outcomes, however, it does assume that collisions between planetesimals are subsonic and cause minimal deformation and phase changes. In RUBBLE all planetesimals are assumed to be gravitational aggregates or rubble piles \citep{Richardson:2002}. When a collision is detected the outcome is determined in the first instance by using a look-up table of pre-calculated rubble-pile collisions. If the largest remnant mass is greater than 80\% of the total colliding mass the collision is deemed ``simple" and the second largest remnant is small, the largest remnant from the look-up table is used as the collision outcome and any remnant mass is relegated to unresolved debris. If the collision is ``complex", the look-up table returns a largest remnant mass that is less than 80\% of the total colliding mass, the planetesimal particles are substituted with rubble-piles (gravitational aggregates of indestructible billiard balls) and the collision is calculated directly within the planet formation evolution simulation. Once the collision has completed any collisional remnants that are less massive than the resolution limit are relegated to unresolved debris. Any collisional remnants more massive than the resolution limit continue to be followed directly by PKDGRAV.

The unresolved debris is distributed in the planet formation region in ten circular annuli. Any mass that is relegated to the unresolved debris is added to the annuli at the instantaneous location of the collision. The unresolved debris is assumed to have circular Keplerian orbits. Resolved planetesimals can accrete unresolved debris in proportion to their geometric cross-section and the eccentricity of their orbit \citep[for details see][]{Leinhardt:2005} as they move through the unresolved debris annuli.

\subsubsection{EDACM: Collision Outcomes} \label{sec:edacm}
\begin{figure}
\includegraphics[width=20pc]{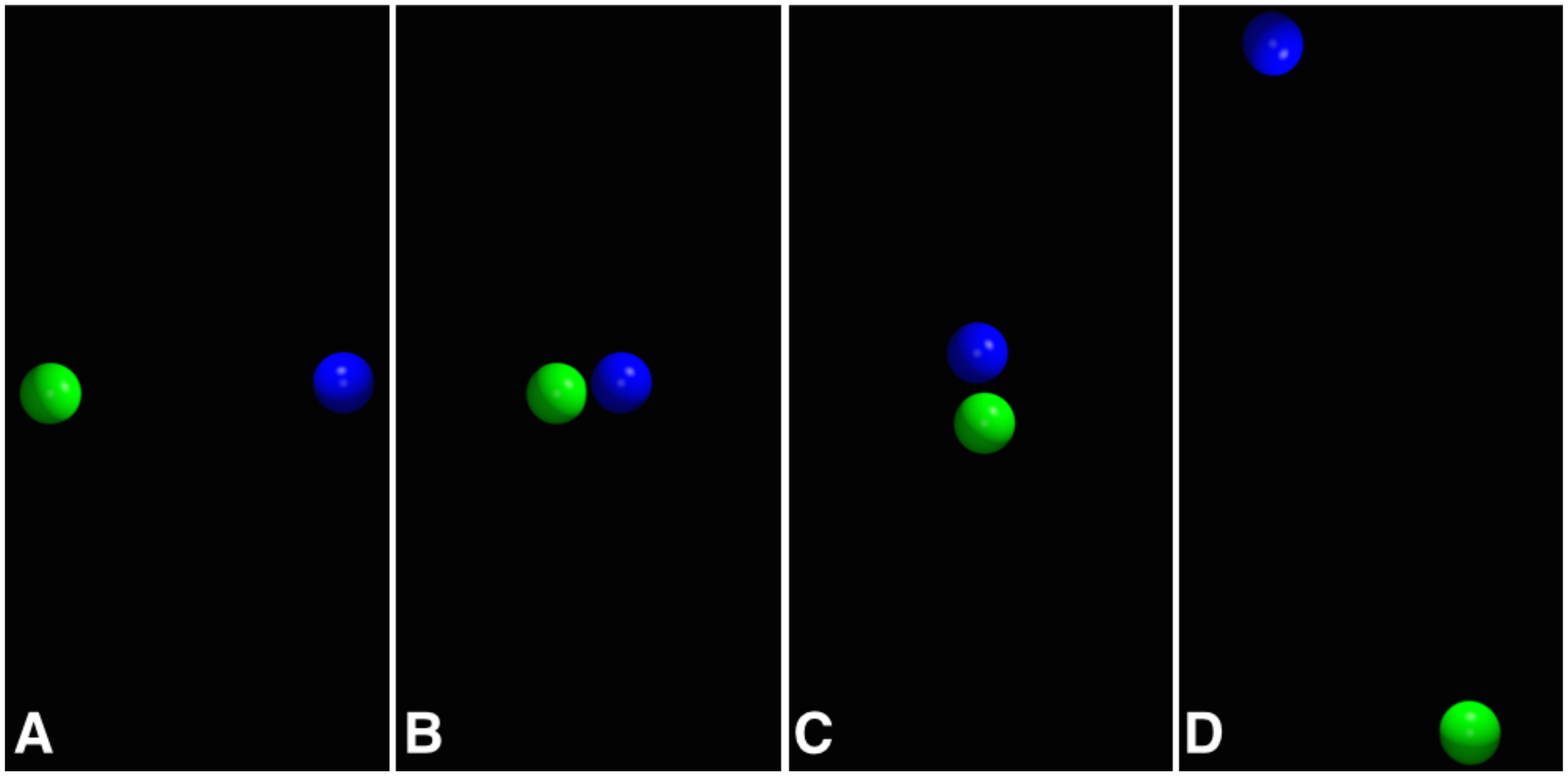}
\caption{Snapshots at $t = 0, 1, 2, 3$ sec for frames A-D showing an example of a ``perfect" hit-and-run collision integrated using the $N$-body code PKDGRAV and the EDACM collision model. The impact shown is a grazing ($b=0.7$), equal-mass (particle radius is 1 km) collision with an impact speed of 50 m s$^{-1}$. }
\label{fig:hitandrun}
\end{figure}

Although the RUBBLE collision model is more accurate than perfect merging, allowing the production of debris, RUBBLE becomes less accurate at high impact speeds because there is no provision for damage or phase changes within the planetesimals. This then limits the versatility of the RUBBLE model. In addition, resolving the more complex collisions is computationally expensive meaning RUBBLE simulations are considerably slower to run than their perfect merging counter parts. In order to rectify some of these draw backs we developed a new empirically derived analytic collision model, EDACM \citep{Leinhardt:2012}, which is both more versatile and computationally faster. 

\citet{Leinhardt:2012} presents a detailed derivation of the EDACM collision model and \citet{Stewart:2012} presents a succinct summary. Thus, here we will only briefly describe the model and instead highlight differences or expansions to the original model. 

In the PKDGRAV implementation of EDACM there are five collision outcome regimes: \\
\\
\noindent 1) \emph{Perfect merging} -- two colliders hit each other and merge into one body. This occurs when the impact speed, $V_i$, is less than the effective escape speed, 
\begin{equation}
V'_{esc} = \sqrt{2GM'_{tot}/(R_p + R_t)},
\end{equation}
where $M'_{tot} = \alpha M_p + M_t$, and $M_p$ is scaled by $\alpha$, the mass fraction of the projectile that geometrically overlaps with the target:
\begin{eqnarray}
\alpha = \left\{\begin{array}{ccc} 1 & : & R_t > b(R_p + R_t) + R_p\\
\rho(\pi R_pl^2 - \pi l^3/3) & : & \mathrm{otherwise},
\end{array}\right. \label{eqn:alpha}
\end{eqnarray}
where $b$ is the impact parameter, and $l/(2R_p)$ is the fraction of the projectile diameter overlapping the target at impact, $R_p, M_p, R_t$, and $M_t$ are the radii and masses of the projectile, and target, respectively. In this collision outcome PKDGRAV replaces the two colliders with a single particle conserving mass and momentum. \\
\\
\noindent 2) \emph{Hit-and-run} -- a collision in which the target and projectile have a grazing encounter, the target looses no mass but the direction of travel is altered. The projectile may or may not suffer erosion depending on the impact energy. 

A hit-and-run event occurs when the impact speed is above $V'_{esc}$ but below the velocity needed for erosion, 
\begin{equation}
V_{erosion} = \sqrt{2 M_{tot} Q_{erosion}/\mu}, 
\end{equation}
where $M_{tot}$ is the total system mass, $\mu$ is the reduced mass, and $Q_{erosion}$ is the specific energy necessary to create a largest remnant that is the mass of the target \citep[see step 4 in][]{Leinhardt:2012} for the parameters of the collision (mass ratio, and impact parameter).  In addition, for the collision outcome to be in the hit-and-run regime the impact must be a grazing event where the impact parameter is greater than the critical impact parameter, 
\begin{equation}
b_{crit} =   \frac{R_t}{R_p+R_t}
\end{equation}
which is defined by \citet{Asphaug:2010} as the impact parameter where the centre of the projectile is tangent to the surface of the target. 

The projectile will remain intact if specific impact energy from the interacting target mass is less than the erosion threshold for the projectile. In PKDGRAV this type of collision is considered a perfect hit-and-run collision and is modelled as an inelastic bouncing event with a normal coefficient of restitution of 0.8. An example of a ``bouncing" hit-and-run is shown in Fig.~\ref{fig:hitandrun}. 

If the impact is more energetic and the specific impact energy exceeds the erosion threshold for the projectile, it will erode. In this case the largest remnant is the original target and the second largest remnant is determined using the universal law which is normally written as
\begin{equation}
M_{lr}/M_{tot} = -0.5(Q_R/Q'^{*}_{RD} - 1) + 0.5,\label{eqn:univ}
\end{equation}
where $M_{lr}$ is the largest remnant, $Q_R$ is the specific impact energy, and $Q'^*_{RD}$ is the critical disruption criterion for the specific collision being considered, or 
\begin{equation}
M_{lr}/M_{tot} = \frac{0.1}{1.8^\eta}(Q_R/Q'^*_{RD})^\eta, \label{eqn:super}
\end{equation}
if $Q_R > 2 Q'^*_{RD}$, meaning the collision is in the super-catastrophic regime, where $\eta = -1.5$ \citep[see appendix][]{Leinhardt:2012}. 
The collisional debris generated is from the erosion or disruption of the projectile only, thus, the $M_{lr}$ is the largest remnant from the projectile disruption  and is actually the second largest remnant in the collision outcome which includes the target. Thus, $M_{tot}$ is simply the mass of the projectile. The initial position, velocity, and mass of the debris field is described in Section \ref{section:debris} below.\\

\begin{figure}
\includegraphics[width=20pc]{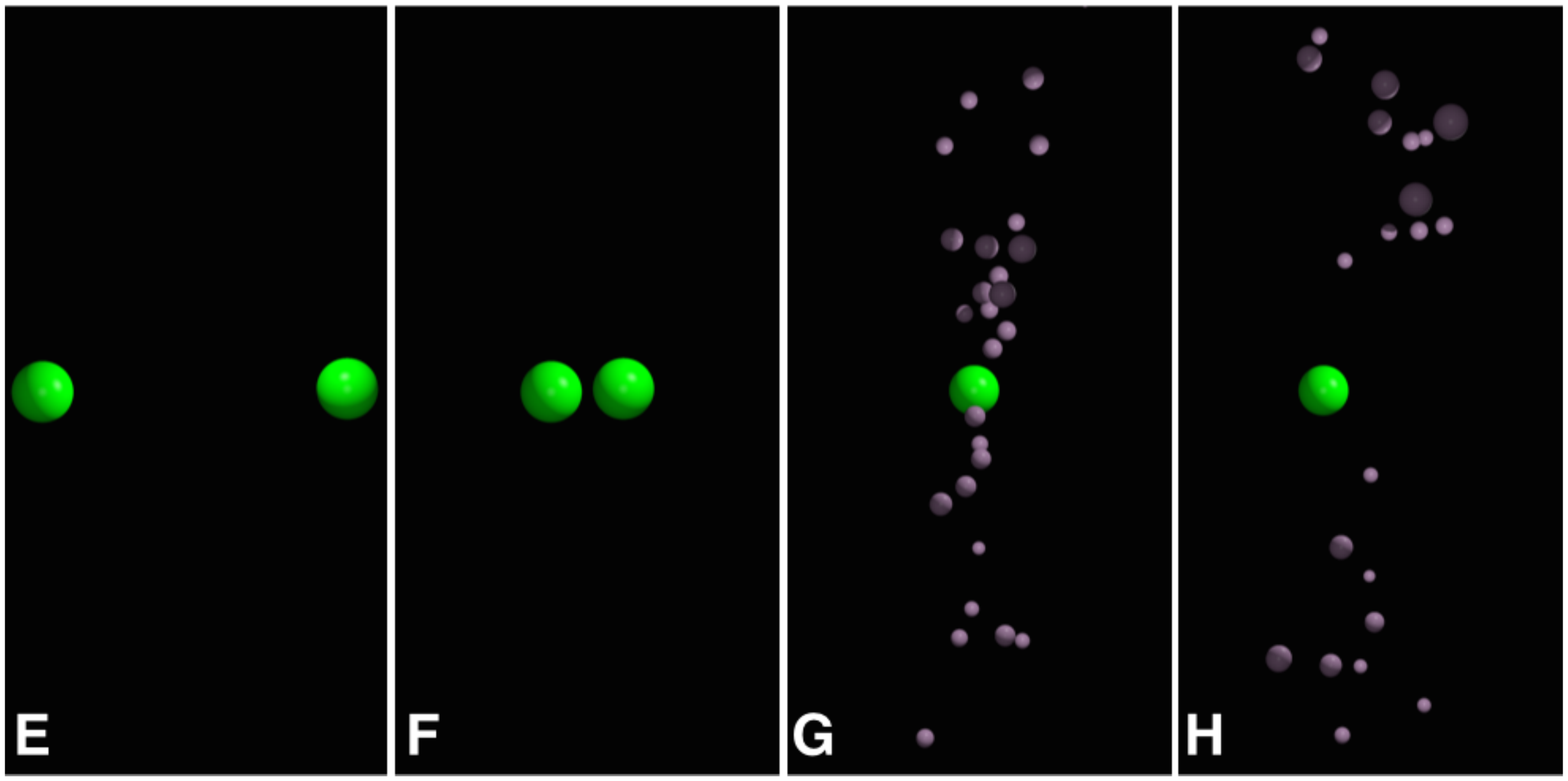}
\caption{Snapshots at $t = 0, 1, 1.5, 2$ sec for frames A-D, respectively, of a partial erosion collision outcome using EDACM implemented PKDGRAV as in Fig.~\ref{fig:erosion}. In this impact the colliders are again equal-mass with a bulk density of 1 g cm$^{-3}$ but have a larger radius ($r=10$ km) than the example shown in Fig.~\ref{fig:hitandrun}, a lower impact parameter ($b=0.2$), and a smaller impact speed ($V_i = 40$ m s$^{-1}$). In frame D the largest remnant, which is 8 km in radius, is shown in green and the debris field is shown in grey. The viewing angle shown is perpendicular to the debris field, meaning this debris field is actually a disk.}
\label{fig:erosion}
\end{figure}

\noindent 3) \emph{Partial accretion} -- two colliders hit each other and the target gains mass. A partial accretion collision resolved with EDACM would look similar to the partial erosion collision shown in Fig.~\ref{fig:erosion} but with the largest remnant more massive than the target. This type of collision event occurs in a non-grazing collision ($b<b_{crit}$) when $V'_{esc}<V_i<V_{erosion}$. In this case $M_{lr}$ is determined using the universal law Eq.~\ref{eqn:univ} and the distribution of the debris field is determined in the same way as it is when a projectile is disrupted in a hit-and-run event (see Section \ref{section:debris} for details).\\
\\
\noindent 4) \emph{Erosion} -- two colliders hit each other and the target loses mass (see Fig.~\ref{fig:erosion}). A partial erosion event occurs when $V_i > V_{erosion}$ and $0.1 M_t < M_{lr} < M_t$. In this regime $M_{lr}$ is still determined by Eq.~\ref{eqn:univ}. In PKDGRAV this means the collision outcome results in the target particle losing mass and the possible generation of a debris cloud of resolved particles. \\
\\
\noindent 5) \emph{Supercatastrophic Disruption} -- two colliders impact each other at high impact speeds, large in comparison to $V'_{esc}$, resulting in $M_{lr} < 0.1 M_t$. The small mass of the largest remnant means there is effectively no separation in the size distribution between $M_{lr}$ and the second largest remnant, $M_{slr}$. In impacts of this speed $M_{lr}$ is determined by the super-catastrophic scaling law Eq.~\ref{eqn:super}. The only difference in implementation within PKDGRAV is the way that the $M_{lr}$ is determined. The debris field is generated in the same way as it is in partial accretion and erosion collision outcomes.\\

\subsubsection{EDACM: Resolution Limit}
Since the EDACM collision model can generate particles as the result of a collision, similar to RUBBLE, we impose a resolution limit, which in most cases is the initial planetesimal size. Collisional remnants with sizes smaller than the resolution limit are treated semi-analytically (see end of Section \ref{sec:rubble}). Any collisional remnants that are larger than the resolution limit are followed directly by PKDGRAV.

 \subsubsection{EDACM: Post-collision Particle Orbits}
The orbits of the particles after collision depend on $b$. In the case of a perfect merging event the one remnant conserves momentum and is placed on the centre of mass orbit. In all other collision outcomes, in which the largest remnant is resolved, $M_{lr}$ is placed at the centre of mass with the centre of mass velocity, ${\bf V_{com}}$, if the collision is head-on ($b=0$). If the impact has a high impact parameter ($b>0.7$) $M_{lr}$ takes on the velocity of the target, ${\bf V_t}$. If the impact was somewhere in-between ($0<b<0.7$) we assume a linear function for the velocity vector, 
\begin{equation}
{\bf V_{lr}} = \frac{{\bf V_t} b + {\bf V_{com}} (0.7-b)}{0.7},
\end{equation}
as suggested in \citet{Leinhardt:2012}.

\subsubsection{EDACM: Debris Field} \label{section:debris}

The size distribution of the debris field (all debris less massive than the largest remnant) is determined by assuming a differential power-law size distribution with an index of -2.85 \citep{Leinhardt:2012}. The remnant masses smaller than and including the second largest remnant were determined by the power-law and conservation of mass (total mass in the debris tail could not exceed $M_{tot}-M_{lr}$).

The velocity distribution of the debris is determined by conserving momentum and ensuring that the total energy of all collisional remnants did not exceed the initial energy of the colliders. Energy is allowed to be lost, however. For simplicity energy equipartition amongst the debris is assumed meaning that the most massive remnant in the debris field will have the slowest speed and the least massive will have the highest speed. Although \citet{Leinhardt:2012} find that the small debris should be found at all speeds the larger debris does travel at the slowest speeds and there was not enough resolution in their simulations to determine how the velocity should be distributed over the remnants more accurately. 
 
 \begin{figure}
 \includegraphics[width=20pc]{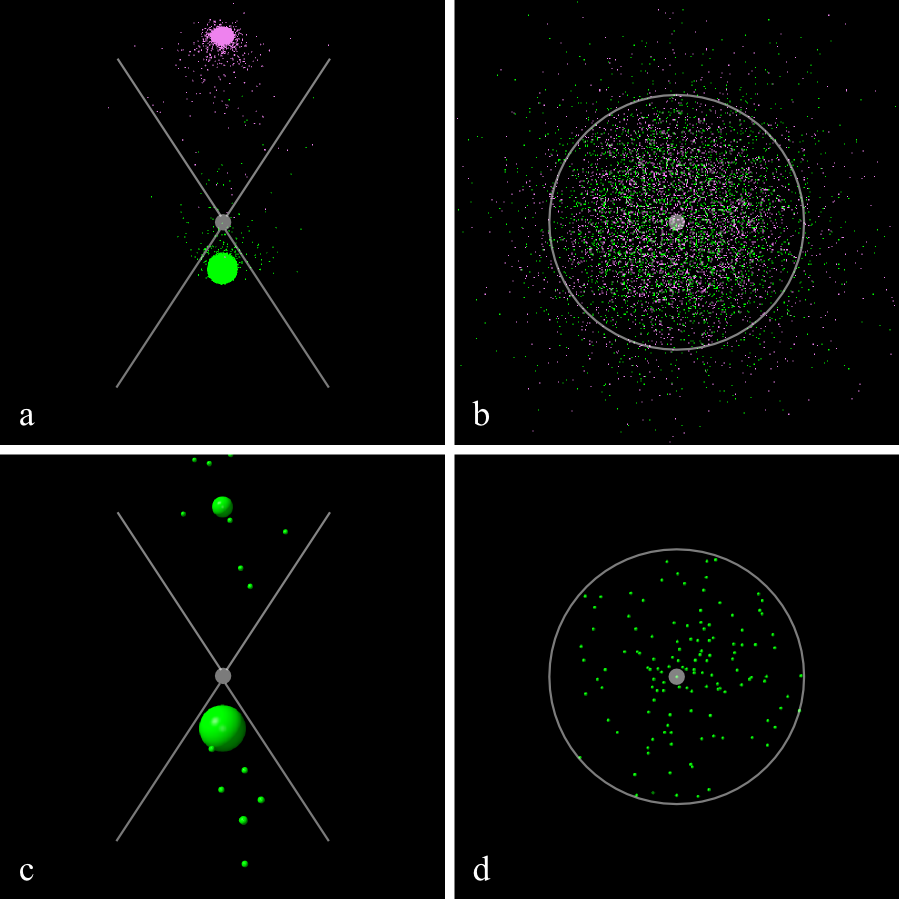}
 \caption{Post-collisional debris viewed face-on looking down on the collision in the $-\bf{u}$ direction (see Fig.~\ref{fig:geometry}). Frames (a) and (b) show the result of a rubble-pile impact. Frames (c) and (d) show the result using the EDACM model. Both are integrated using PKDGRAV. The impact shown in (a) and (c) is a hit-and-run projectile disrupt ($v_i = 50$ m s$^{-1}$, oblique impact: b=0.9), between a km-sized target and a projectile a quarter of the mass. The impact shown in (b) and (d) involves the same projectile and target but is head-on and super-catastrophic ($v_i = 38$ m s$^{-1}$, b = 0.0).}
 \label{fig:disp}
 \end{figure}
 
The spatial distribution of the particles in the debris field is complex. In some impacts debris forms a disk around the largest remnant but in many collisions the debris forms two jet-like features (Fig.~\ref{fig:disp}). In some instances these ``jets" are imbalanced and remove residual momentum from the largest remnant and in other cases these ``jets" are symmetric and account for zero net momentum in the collision outcome. In order to determine the most accurate placement for the resolved fragments we analysed the debris distributions of the several hundred rubble-pile simulations used to create the EDACM collision model presented in \citet{Leinhardt:2012} and derived a series of rules for post-collision debris placement. Two very different collision debris fields from isolated rubble-pile collisions modelled with PKDGRAV are shown in Fig.~\ref{fig:disp}a-b. The EDACM generated version of these collisions is able to capture the general character of the debris field as shown in frames c-d. The details of the model implemented in EDACM are given below.

Two angles are used to describe the distribution of fragment material $\theta$ ($0^\circ \le \theta < 360^\circ$), which determines the direction of the debris swarms and $\phi$ ($-90^\circ \le \phi < 90^\circ$), which determines the degree of ``jetiness" or ``diskiness". The fragment probability distribution (a bi-Gaussian distribution in $\theta$ and a Gaussian distribution in $\phi$),
\begin{equation}
P_{\theta \phi} = A_{\theta_1}e^{\frac{-(\theta-\theta_1)^2}{2\sigma_\theta^2}}e^{\frac{-\phi^2}{2\sigma_\phi^2}} + A_{\theta_2}e^{\frac{-(\theta-\theta_2)^2}{2\sigma_\theta^2}}e^{\frac{-\phi^2}{2\sigma_\phi^2}},
\end{equation}
is defined by two peaks in $\theta$ located at $\theta_1$, $\theta_2$, and one at $\phi=0^\circ$ with thicknesses represented by the standard deviations $\sigma_\theta$ and $\sigma_\phi$ and relative amplitudes in $\theta$ denoted as $A_{\theta_1}$, $A_{\theta_2}$. 

\begin{figure}
\includegraphics[width=20pc]{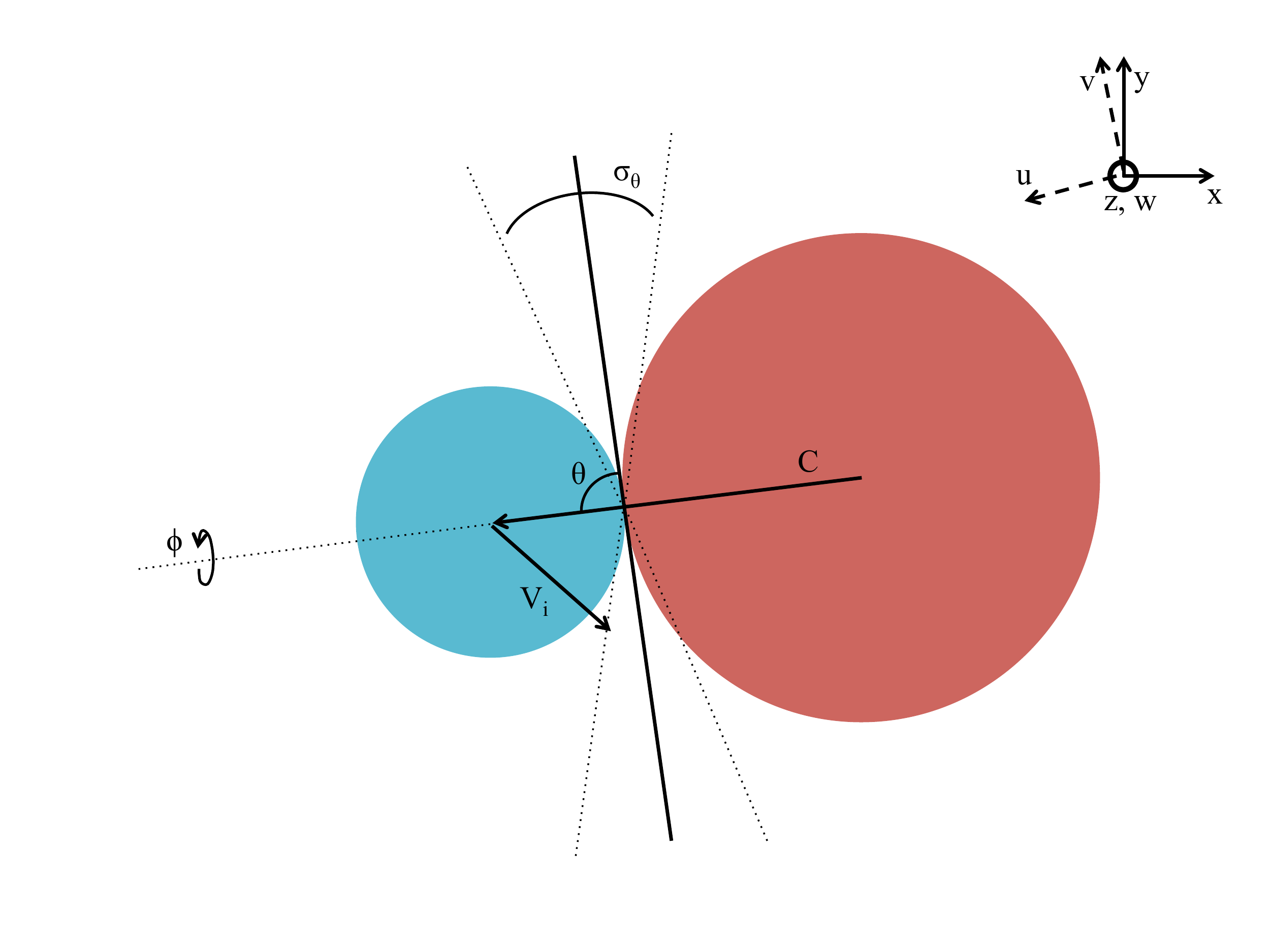}
\caption{Cartoon of the geometry of an impact between a projectile (blue) and a target (red). The angle $\theta$ indicates collisional debris axis. The value of $\phi$ indicates whether the debris is distributed in two jets or in a complete disk around the impact site.}
\label{fig:geometry}
\end{figure}

In order to determine the locations of all of the debris material let us first define a convenient collisional coordinate system $\bf{u,v,w}$. The first basis vector, $\bf{u}$ is in the direction of the center joining vector, ${\bf C} = {\bf R_p} - {\bf R_t}$, where $\bf R_p$ and $\bf R_t$ are the positions of the projectile and target respectively (see Fig.~\ref{fig:geometry}), thus, 
\begin{equation}
\bf u = C/|{C}|.
\end{equation} 
The direction of the $\bf w$ basis vector is perpendicular to the impact velocity vector, $\bf V_i$ and $\bf C$
\begin{equation}
\bf w =  V_i/|V_i| \times u
\end{equation}
and 
\begin{equation}
\bf v = u \times w.
\end{equation}
Thus, the impact vector, $V_i$, is in the $u-v$ plane. 

The debris axis is in the $u-v$ plane and is defined by two peaks 
\begin{eqnarray}
\theta_1 &=& \left\{\begin{array}{ccc} 70\,b & : & \,b \ge b_{crit}\\
\frac{70\,b_{crit} - 90}{b_{crit}}\,b + 90 & : & \,b < b_{crit}
\end{array}\right.\\
\theta_2 &=& \theta_1 +180,
\end{eqnarray}
with $\theta = 0^\circ$ in the $\bf u$ direction.
If the collision was a non-grazing, head-on collision ($b=0$) the debris axis would be distributed about $\theta_1 = 90^\circ$ and $\theta_2 = 270^\circ$. Debris moves away from the impact site in two directions $180^\circ$ apart with a narrow width that is relatively independent of collision parameters. Thus, the standard deviation of the debris distribution in $\theta$ is set as a constant, 
\begin{equation}
\sigma_\theta = 10^\circ.
\end{equation}

\begin{figure*}
\includegraphics[height=21pc]{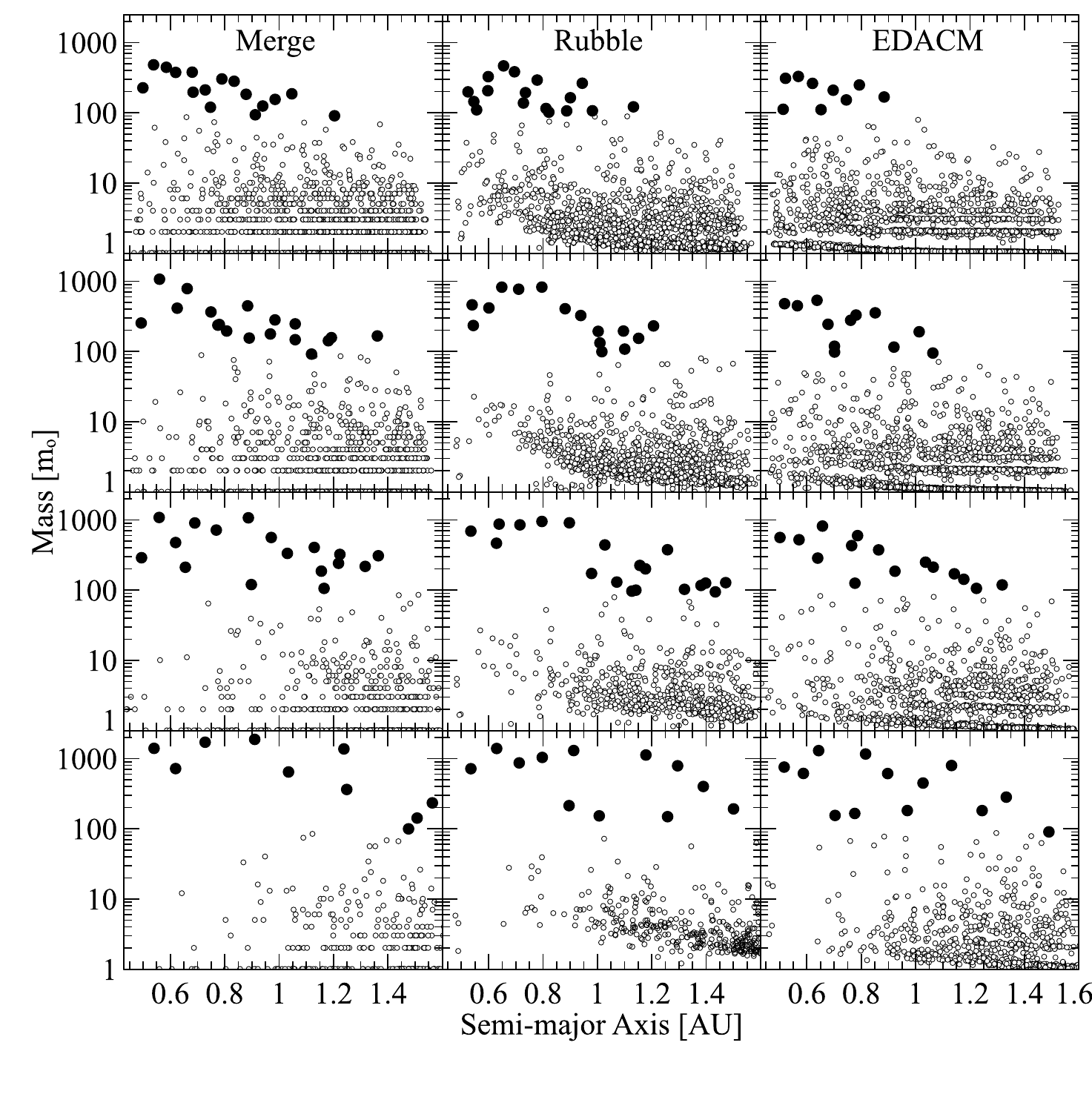}\includegraphics[height=21pc]{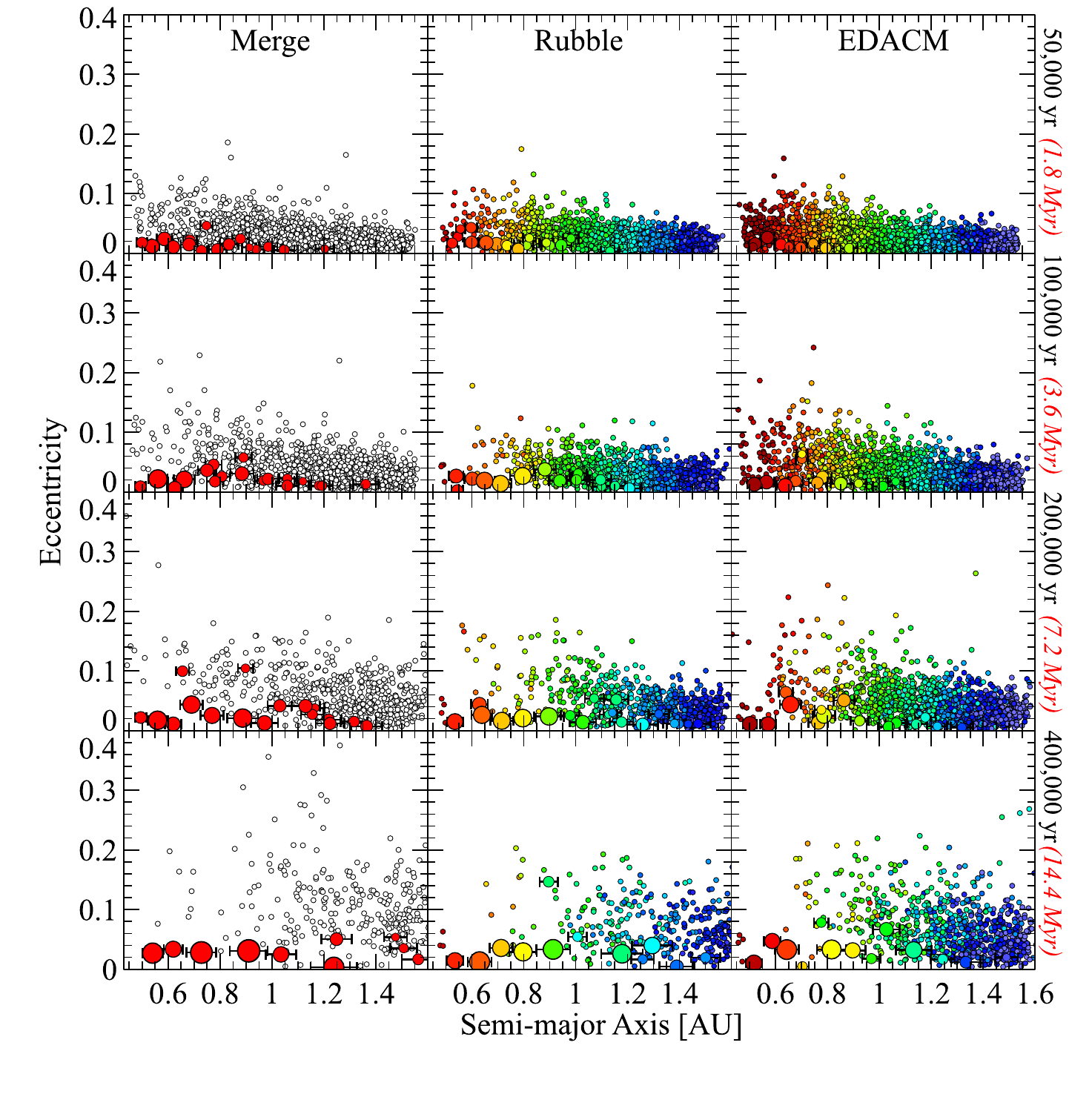} 
\caption{Evolution of mass and eccentricity of resolved particles (planetesimals and protoplanets) versus semi-major axis for three collision models: merging, RUBBLE, and EDACM. Protoplanets, objects that have reached isolation mass (100 times initial mass), are indicated with black filled circles (left) and coloured circles with error bars (right). Planetesimals, objects that are directly resolved but have not reached isolation mass, are shown as small open or coloured circles. Time increases from top to bottom. The simulation time for each row is indicated in black on far right. The red italicised time is the estimated \emph{effective} time for un-inflated planetesimals with expansion parameter of one. Right - error bars show gravitational influence of protoplanets and extend 5 Hill radii on either side. The size of the protoplanets is scaled by radius. Rubble and EDACM collision model columns also have colour derived from composition histograms associated with each particle.}
\label{fig:avse_all}
\end{figure*}

The relative amplitudes of the two peaks in $\theta$ depend strongly on impact parameter $b$ and projectile to target mass ratio, $\mu$, such that,
\begin{equation}
\frac{A_{\theta_1}}{A_{\theta_2}} = 1 - b(1 - \mu).
\end{equation}
Thus, in all equal-mass collisions the debris is symmetric and independent of $b$, however, for unequal mass impacts the debris can carry momentum away from the impact site.

The peak of $\phi$ distribution, $\phi = 0^\circ$, is in the $\bf v$ direction. The width in $\phi$ depends most prominently on the fraction of impacting mass,
\begin{equation}
\sigma_\phi = 145\,\alpha^\circ,
\end{equation}
where $\alpha$ is the projectile mass fraction involved in the collision (as defined in Eq.~\ref{eqn:alpha}). Note that the $\phi$ distribution can be broad enough that the two debris ``jets" actually touch creating a full disk (see Fig.~\ref{fig:disp}d).

\section{Results}

The general evolution of the planetesimals is independent of the collision model (Fig.~\ref{fig:avse_all}) because most collisions in the relatively calm evolution scenario used in this work are accretion dominated and in addition, the velocity dispersion is dominated by the gravity of the largest planetesimal bodies. This results in the largest planetesimals following a similar growth history. The similarity seen here between the three collision models may not be seen in systems that undergo an energetic dynamical shake-up and/or have a large perturber \citep[for example, the Grand Tack scenario,][]{Walsh:2011}. In any case, in the perturber-free scenario modelled here subtle differences amongst the different models are detectable in the evolution of the background planetesimal population which suffers more energetic collisions that result in fragmentation. As the largest embryos grow the RMS velocity of the background increases. The EDACM model allows fragmentation and provides both directional and velocity information about the collision remnants, thus we can make a prediction of observable dust (see \ref{sec:dust}). 

\subsection{Planetesimal Evolution}

\begin{figure*}
\includegraphics[width=42pc]{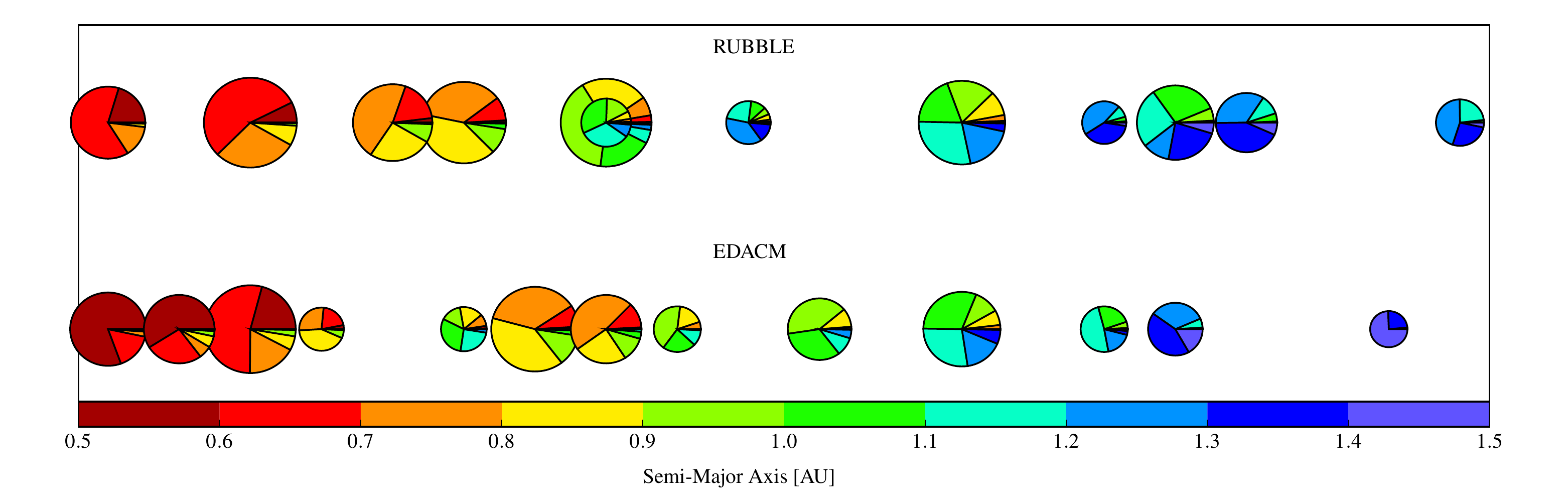}
\caption{Composition pie plots of each protoplanet from the RUBBLE and EDACM runs. The protoplanets are shown at their instantaneous semi-major axis. The size of the protoplanet is proportional to the object's radius. Each wedge of a pie indicates the fraction of mass from the designated radial region shown by the colour bar at the bottom of the figure. }
\label{fig:mixing}
\end{figure*}

Figure \ref{fig:avse_all} shows snapshots of planetesimal evolution and embryo growth for three collision models: perfect merging, RUBBLE and EDACM. The left hand panels show mass versus semi-major axis and right show eccentricity versus semi-major axis. The larger data points in these frames are used to indicate planetesimals that have reached isolation mass \citep[$\approx 0.1 \mathrm{M_\oplus} \approx 100 \mathrm{m}_\mathrm{o}$,][]{Leinhardt:2005}. In all simulations the embryo population has begun to separate from the background population at small semi-major axis by 100,000 yrs (simulation time). By the end of the simulations all protoplanets are approaching a similar mass indicating a transition from runaway to oligarchic growth \citep{Kokubo:2002,Leinhardt:2005}. 

As would be naively expected the perfect merging simulation evolves the most quickly because it contains no fragmentation, then RUBBLE, and finally the EDACM simulation. 
In addition, there is no significant change to the growth modes due to different collision models. The only noticeable difference between the models in Fig.~\ref{fig:avse_all} is a slight increase in the evolution timescale due to less efficient growth when fragmentation is included. This may get more pronounced with evolution time because the average impact speeds increase and more of the collisions are erosive. None the less, once the difference in evolution timescale is taken into account the three collision models qualitatively agree in eccentricity and inclination distributions, number, and mass of protoplanets. 

The RUBBLE and EDACM collision models allow the composition of all particles to be tracked. The composition of the protoplanets (shown in Fig.~\ref{fig:mixing} as pie plots) show some localised mixing. The pie plots are based on the composition histograms of each protoplanet which track where the mass accreted by each particle comes from. In general, most of the mass in each protoplanet comes from $\pm 0.1$ AU of the protoplanet's current semi-major axis. This can also be seen by examining the colours of the protoplanets in Fig.~\ref{fig:avse_all} which indicate a mass-weighted average of the protoplanets composition (a mean of the pie plot). There is no significant difference in the degree of mixing between the two collision models.

\begin{figure}
\includegraphics[height=20.5pc]{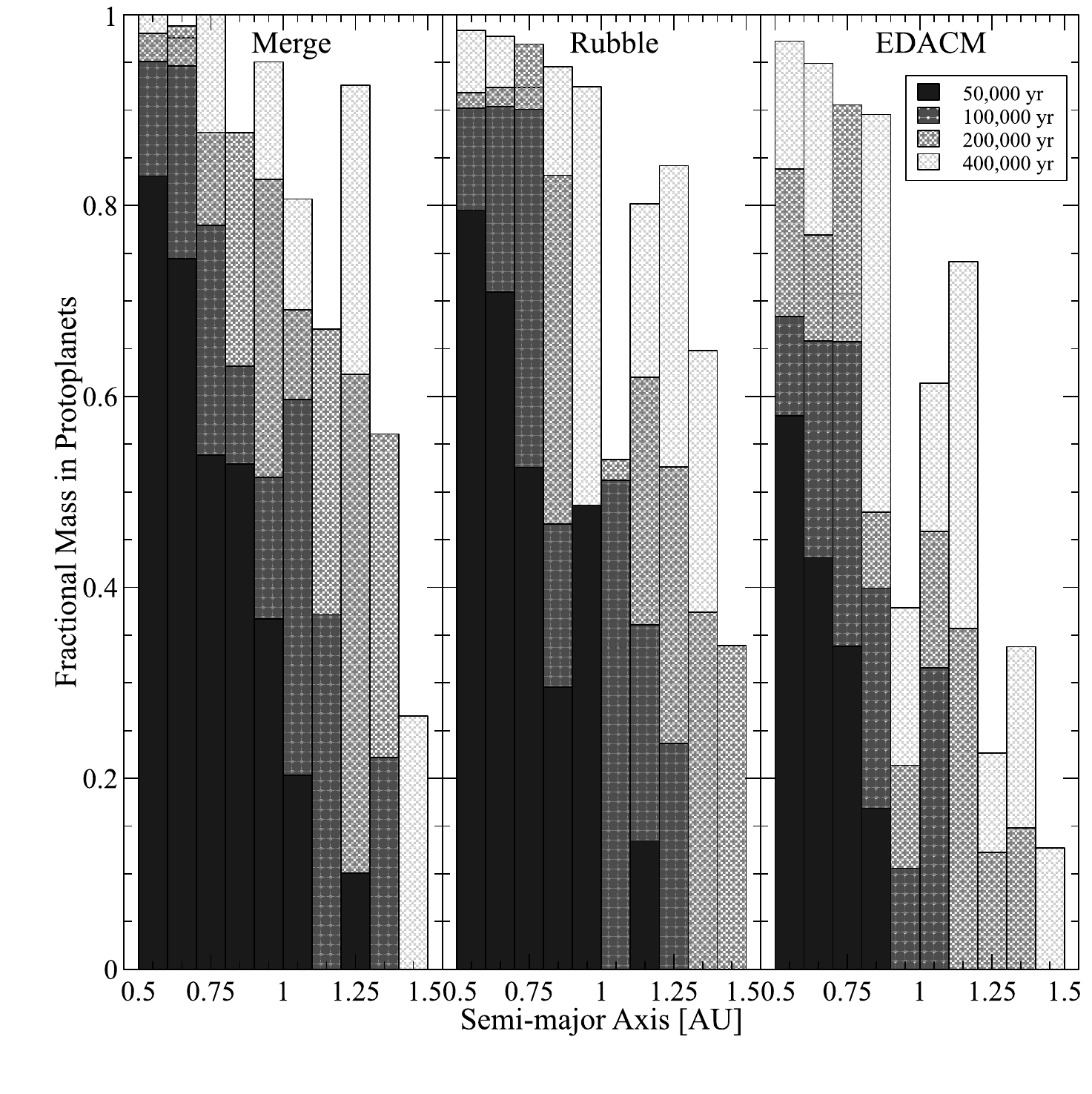}
\caption{Fractional mass in protoplanets as a function of radial distance for the three collision models. The shaded histograms show the mass in protoplanets over a radial bin of 0.1 au normalised by the total mass in that bin. Time evolution is indicated by the different shades of grey increasing with time from dark to light.}
\label{fig:masshist}

\end{figure}

In all simulations, no matter what collision model is used, we find that the background planetesimal population is efficiently cleared out by the embryos. There is no significant period of time over the 1AU radial annulus in any simulation in which half the mass is in small planetesimals and half the mass is in large protoplanets (Fig.~\ref{fig:masshist}). Moreover, the difference in evolution timescale in the inner region of the annulus and the outer region of the annulus is noticeable even within 1 AU. The inner region of the annulus (inward of 0.8 AU) has been all but entirely cleared of planetesimals by the end of the simulations where as a considerable fraction of the mass if not the majority is in the planetesimal population in the outer regions of the annulus. These results bring into question arguments made in previous work such as, \citet{Goldreich:2004}, that a significant mass can be stored in the planetesimal population and thus reduce the orbital eccentricities of the oligarchics via dynamical friction at late times.

\subsection{Collisions}

\begin{figure}
\includegraphics[width=20.5pc]{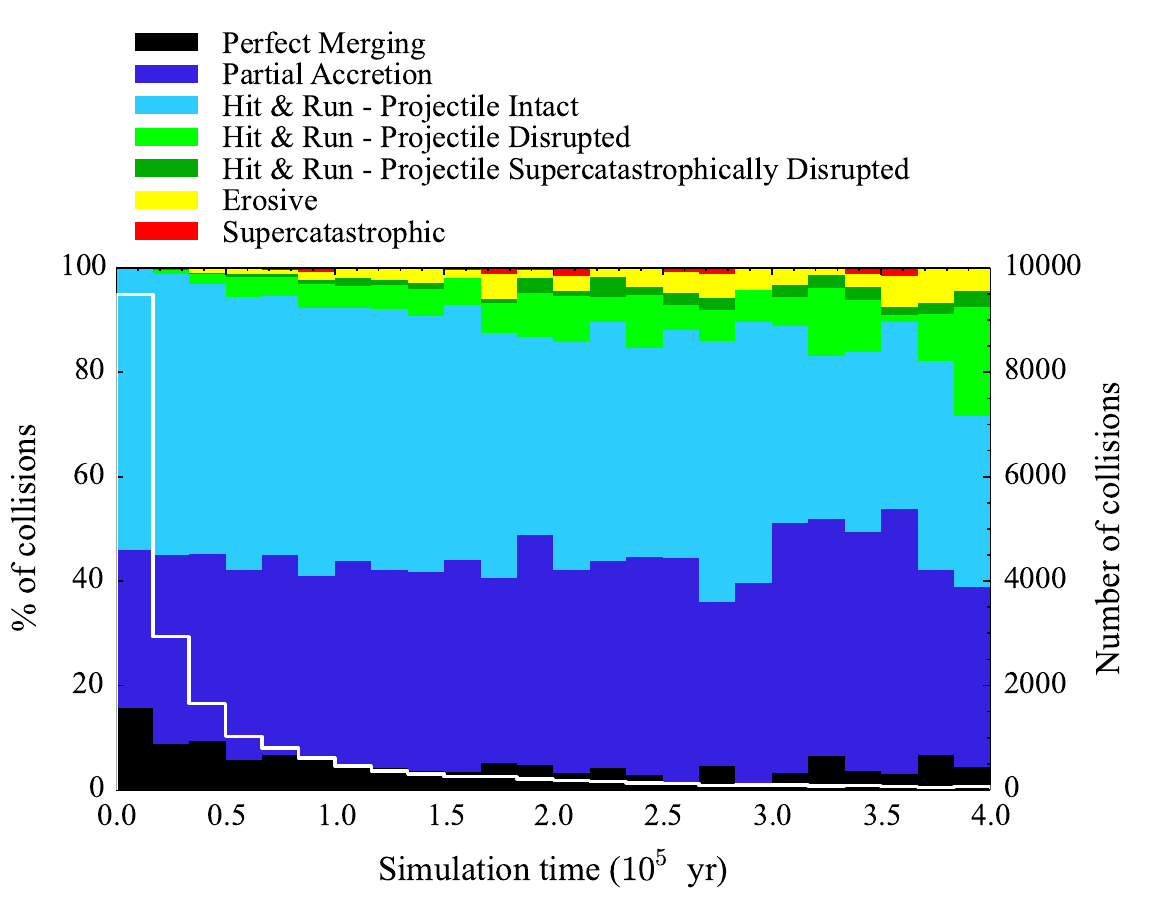}
\caption{Coloured histogram shows the percentage of planetesimal collision type per time bin for the extent of the simulation (left y-axis). Each time bin has a width of $2\times10^5$ yr in simulation time. The white outlined histogram indicates the number of collisions as a function of time (right y-axis).}
\label{fig:collisionhist}
\end{figure}

Although most of the collisions in RUBBLE and EDACM simulations are accretion dominated very few of the collisions in the EDACM collision model actually result in perfect merging (see black sections in each bar of Fig.~\ref{fig:collisionhist}), instead the collision outcomes are divided between partial accretion events (dark blue) and some type of hit-and-run (light blue -- projectile intact, light green -- projectile disrupted, and dark green -- projectile catastrophically disrupted). The fraction of partial accretion events remain relatively constant just above $40\%$ while the fraction of hit-and-run events with an intact projectile (light blue) begins to decrease with time. This is due to the increase in the average impact speed as the largest bodies grow and gravitationally stir the rest of the objects. With an increase in speed even impacts with too much angular momentum to result in an accretion or merging event deliver enough energy to disrupt or partially disrupt the projectile. In addition to the decrease in ``perfect" hit-and-run events with time erosive events also begin to increase in number (yellow and red). Again this is due to the increased average speed of collisions due to gravitational stirring from growing planetary embryos. However, it is clear that the vast majority of all collisions are non-erosive (for the target) during all phases of evolution. 

The number of collisions decreases exponentially with time (white outlined histogram) as runaway growth quickly reduces the number of particles available for collisions. As a result, the collision type histogram becomes more noisy with time as the low number of collisions reduces the statistical accuracy of each late time bin. The collisional breakdown shown in Fig.~\ref{fig:collisionhist} shows slightly more hit-and-run projectile intact and slightly less catastrophic disruption events than found in \citet{Bonsor:2015} which investigates the formation of the Earth using similar numerical simulations. 
The difference in these results is due to a difference in initial size distribution and resolution limit and not to any subtle difference in the collision model. 

Figure \ref{fig:avse_all} shows that varying the collision model does not significantly effect the growth phases of runaway or oligarchic growth due to the dominance of bouncing and accretion dominant collisions (Fig.~\ref{fig:collisionhist}). However, including a realistic fragmentation model does allow us to make observational predictions which would not otherwise be possible.


\begin{figure}
\includegraphics[height=18pc]{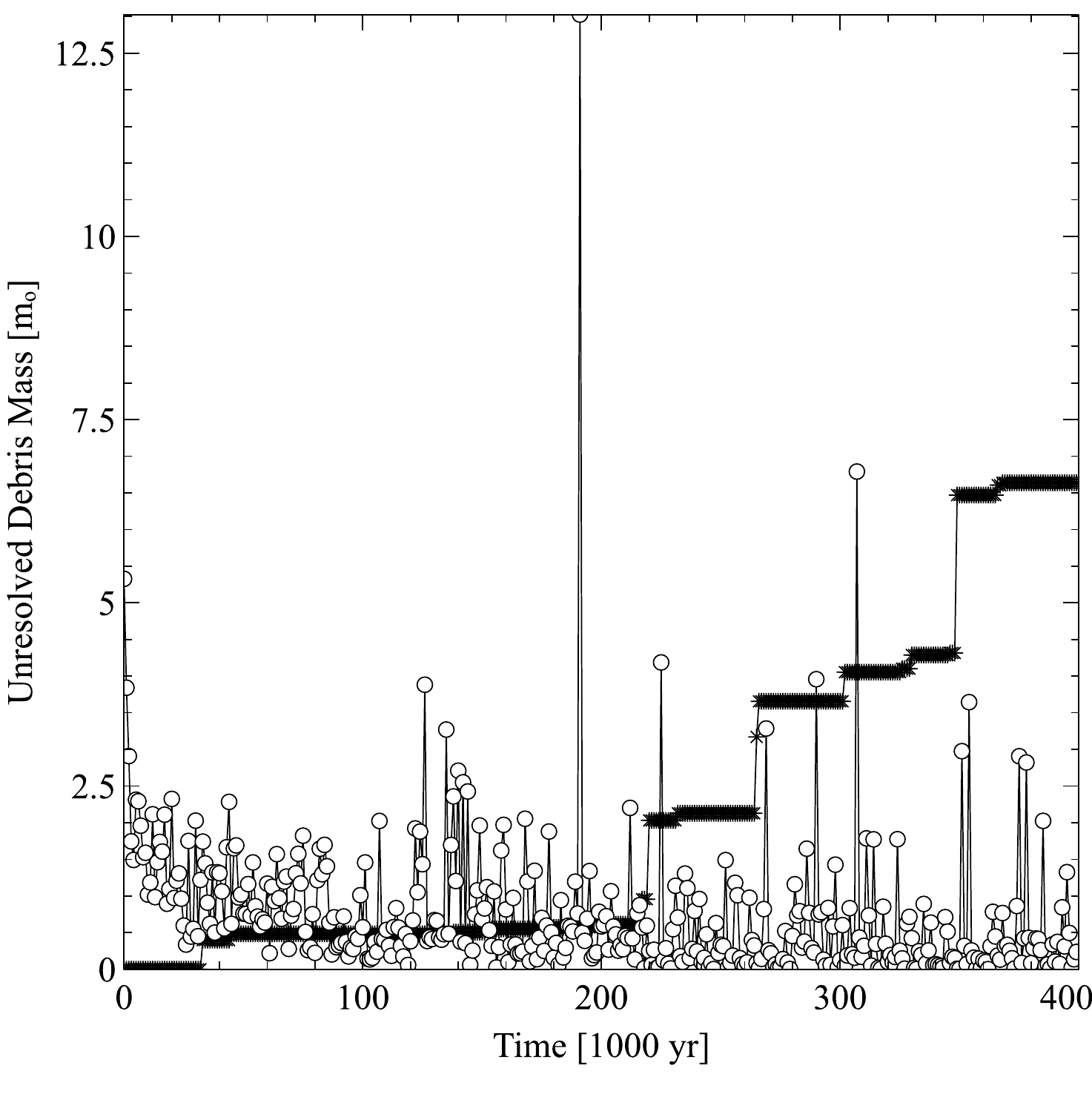}
\caption{Evolution of unresolved debris in EDACM simulation. The total mass of unresolved debris in units of initial planetesimal mass $m_o$ is show with open circles. Unresolved debris outside of the integration annulus, which lies between 0.5 and 1.5 AU, is shown with black crosses and is not available to be accreted by resolved planetesimals. 
}
\label{fig:unresolved}
\end{figure}

\section{Discussion}

Although the direct numerical simulations presented here have a fixed resolution limit of $\sim$ 500 km we can use the orbital information from the resolved planetesimals along with the mass in unresolved material (Fig.~\ref{fig:unresolved}) to help us make predictions of the location and intensity of dust visible during the formation of planetary embryos. In order to do this we follow the method described in \citet{Dobinson:2013} which is briefly summarised in the section below. 

\subsection{Synthetic Dust Images}\label{sec:dust}

\begin{figure*}
\includegraphics[width=40pc]{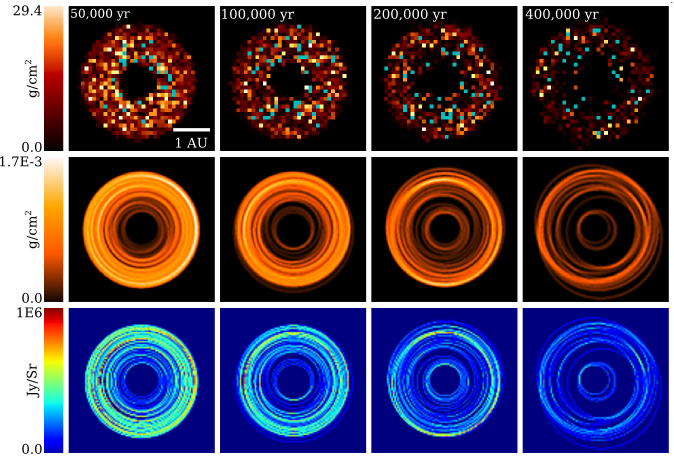}
\caption{Planetesimal evolution during runaway and oligarchic growth shown in planetesimal surface density (top row), instantaneous predicted collisional dust surface density (middle row), and as synthetic images of instantaneous dust (bottom row). Saturated pixels are shown in the planetesimal surface density images in cyan. The synthetic images were made using the radiative transfer package RADMC3D at a wavelength of 850 $\mu$m. Dust species are modeled as amorphous carbon and silicates, both species have sizes of 0.1 and 0.631 $\mu$m and relative abundances of 0.2 and 0.8, respectively. Larger dust from 1 to 1000 $\mu$m is modeled using simple Mie scattering spheres.}
\label{fig:resolved_dust}
\end{figure*}

The first step in creating a synthetic image of dust is to determine the dust surface density. Predicting the dust surface density beyond the resolution limit of a numerical simulation has also been used in work investigating the formation and detectability of debris disks \citep{Booth:2009}. However, in that case the authors assumed that the resolved planetesimals would never grow and provided only source material for dust. Thus, in the case of debris disk formation the planetesimal mass could be used as a direct tracer for dust, namely, the more massive the planetesimal, the more dust generated in a collisional cascade. In the planet formation environment presented in this work the planetesimals are not in any form of steady state, instead they are growing and evolving due to collisions with each other. Therefore, we would expect dust to trace dynamical activity more directly than mass. In addition, we will also assume that all collisions produce some dust even if the collision between the parent planetesimals results in a perfect merging event. The total dust mass is determined by the mass flux of resolved planetesimals on crossing orbits (see resolved planetesimal surface density in first row of Fig.~\ref{fig:resolved_dust}). The mass in dust generated by predicted collisions with each planetesimal is smeared over the orbit of the target planetesimal with mass distributed in equal time chunks, thus, highly eccentric orbits are not uniformly bright. The total mass in dust over the entire region is then scaled to the average total mass in unresolved debris close in time to each time step plotted (see Fig.~\ref{fig:resolved_dust} second row). The instantaneous dust mass is averaged to reduce the noise in the dust surface density due to impact events at or near a given time step, which can be seen in the spikes and valleys in Fig.~\ref{fig:unresolved}.

A telescope would only observe a fraction of this total dust mass - namely, dust of a size similar to the observed wavelength. In addition, the brightness of the dust will depend strongly on the proximity of the dust to the central star, which is assumed to be solar. Taking this into account the third row of Fig.~\ref{fig:resolved_dust} shows synthetic images produced using the radiative transfer code RADMC3D \citep{Dullemond:2012} at an ALMA observing wavelength of 850 $\mu$m, assuming a maximum of one hundredth of the total unresolved debris mass at each time step ends up in dust between 0.1 and 1000 $\mu$m (roughly estimated from assuming a power-law size distribution with a slope of $-3.5$), where the lower limit on the dust size is the blow-out radius and the upper limit on the dust size is determined by the largest dust particle that significantly contributes to the sub-mm flux. The resulting images show a disk with a maximum flux of about $1\times 10^6$ Jy/sr that both dims and becomes more ring-like as planetesimal evolution and protoplanet formation occur in ernest in the inner terrestrial region. 

What would be required to resolve such structures with ALMA? If we assume that the average brightness of $5\times 10^5$ Jy/sr and that we need to resolve disk structure on the scale of 0.2 AU at 10 pc then our beamsize is 0".02 which gives us a solid angle of $1.06\times10^{-14}$ sr. This makes the flux per beam or resolution element $5.3\times10^{-9}$ Jy. Observing using the full ALMA array at 353 GHz (850 $\mu$m) gives a sensitivity equal to the flux in $5.25\times 10^5$ days. Obviously not practical, however, if the flux scaled with mass and the disk were 1000 times more massive such a disk would be detectable with a 12 hour observation. A disk with $1000$ times more mass may not be that extreme. The protoplanetary disk shown in this work is only $3\times3$ AU$^2$ which is over an order of magnitude smaller than the current extent of our own Solar System and almost two orders of magnitude smaller than the young protoplanetary disk imaged by ALMA around HL Tau. The disk investigated in this work assumed a minimum mass solar nebula which is most likely an underestimate for our own solar system. The diversity of extrasolar planets and protoplanetary disks certainly suggest that accretion disks could be much more massive. In addition, the scenario presented in this paper involved no external perturbers such as a nearby giant planet or binary companion, which may significantly enhance planetesimal collision velocities, dust production, and increase the possibility of observational detection (see Dobinson et al.~sub, Carter et al.~in prep). Thus, if we observed the combined flux from a dust disk that contained 1000 times more dust mass it would be marginally detectable with ALMA\footnote{almascience.eso.org/proposing/sensitivity-calculator}.

\section{Conclusion}

Large swaths of planet formation are currently unobservable in a direct sense due to lack of emission from the evolving planetesimals. However, indirect observation of planetesimal growth may be possible in the near future by observing collisionally generated dust with cutting edge observatories. In this paper we present our predictions for these observations using the efficient N-body code PKDGRAV to model runaway and oligarchic growth coupled with EDACM, an empirically derived collision model that allows multiple collision outcomes including perfect merging, accretion, and erosion. We find that planetesimal growth and evolution could be indirectly observable with a fully functioning ALMA-like telescope via collisionally generated dust even in a calm system if the total dust production was large enough. 

\section{Acknowledgements}
The authors acknowledge support from the Science and Technology Facilities Council and the Natural Environment Research Council. The authors also wish to thank John Chambers for a thoughtful review of this manuscript.

\bibliography{GeneralBib}
\end{document}